\documentclass[%
 aip,
 amsmath,amssymb,
preprint,%
]{revtex4-1}

\usepackage{graphicx}
\usepackage{dcolumn}
\usepackage{bm}

\usepackage[utf8]{inputenc}
\usepackage[T1]{fontenc}
\usepackage{mathptmx}
\usepackage{etoolbox}
\usepackage{braket}
\usepackage{comment}

\usepackage{xcolor}

\begin{document}

\title{Configuration interaction extension of AGP for incorporating inter-geminal correlations}
\author{Airi Kawasaki}
\email{a_kawasaki@gunma-u.ac.jp.}
\affiliation{Division of Electronics and Mechanical Engineering, Graduate School of Science and Technology, Gunma University, 1-5-1 Tenjin-cho, Kiryu-shi, Gunma 376-8515, Japan}
\author{Fei Gao}%
\affiliation{Department of Physics and Astronomy, Rice University, Houston, Texas 77005-1892, United States}%
\author{Gustavo E. Scuseria}
\affiliation{Department of Physics and Astronomy, Rice University, Houston, Texas 77005-1892, United States}%
\affiliation{Department of Chemistry, Rice University, Houston, Texas 77005-1892, United States}%

\date{\today}

\begin{abstract}
In this paper, we develop a class of antisymmetrized geminal power configuration interaction (AGP-CI) wave functions that extend the AGP framework by incorporating inter-geminal correlations through a CI expansion. To make these wavefunctions computationally tractable, we evaluate them by rewriting the AGP-CI ansatz as a linear combination of AGPs (LC-AGP), for which overlaps and Hamiltonian matrix elements can be computed with standard AGP machinery. Motivated by border-rank decompositions, we further reorganize this ansatz into a compact linear combination of AGPs depending on a small deformation parameter $\tau$, which controls how closely the truncated expansion approximates the full AGP-CI state. Benchmark applications to the Hubbard model and to the small molecules H$_2$O and N$_2$ demonstrate that the proposed wavefunctions achieve consistently high accuracy and outperform the LC-AGP, particularly for systems with more electrons and in strongly correlated regimes.

\end{abstract}

\maketitle

\section{Introduction}

The accurate description of electron correlation remains a central challenge in quantum chemistry \cite{aszabo82:qchem, pople1999nobel}. Beyond the independent particle approximation to incorporate electron correlation, various computational methods have been developed, including many-body perturbation theory \cite{kelly1964many}, coupled-cluster theory (CC) \cite{bartlett2007coupled} and configuration interaction (CI)  \cite{shavitt1998history} approaches. However, these approaches are fundamentally constructed upon the one-body electron approximation, and their performance deteriorates as the electron correlation becomes stronger.

For precise calculations in systems exhibiting significant electron correlation, that is, strongly correlated electron systems, methods based on geminals \cite{shull1959natural, 1965JMP.....6.1425C, tecmer2022geminal}, which are constructed from correlated electron pairs, have recently attracted considerable attention.

A geminal can be defined as
\begin{eqnarray}
 \hat{F} = \sum_{ab}^{2m} F_{ab} \hat{c}^{\dag}_{a}\hat{c}^{\dag}_{b}.
\end{eqnarray}
where $2m$ is the number of spin orbitals and $F$ is an antisymmetric matrix.
The simplest wavefunction constructed from this geminal is the antisymmetrized geminal power (AGP) wavefunction \cite{1965JMP.....6.1425C, 10.1063/1.1697193}, which can be written as follows.
\begin{eqnarray}
\ket{\Psi_{\mathrm{AGP}} } = \hat{F}^n \ket{-} \equiv \ket{F}
\end{eqnarray}
where $2n$ is the number of electrons and $\ket{-}$ denotes the vacuum state. For notational simplicity, we omit the normalization factor $1/n!$.
The AGP wavefunction constructed from identical geminal products can capture correlations within electron pairs, but correlations between pairs are treated only at the mean-field level. A simple strategy to include inter-geminal correlations is to construct a wavefunction as a linear combination of AGP wavefunctions using distinct geminals \cite{2015PhRvA..91f2504U, uemura2019antisymmetrized, dutta2020geminal, dutta2021construction}, referred to as LC-AGP, which can be written as follows.
\begin{eqnarray}
\ket{\Psi_{\mathrm{LC-AGP}} } &=&  \sum_k^K \left( \hat{F}^{(k)} \right)^n \ket{-} = \sum_k^K\ket{F^{(k)}}
\end{eqnarray}
Here, $k$ denotes the index of the geminal type and $K$ is the total number of geminal types (the number of AGP terms) used. The coefficients associated with each term in the LC-AGP wavefunction are omitted, as they can be absorbed into the matrix $F$.
The overlaps of the AGP and LC-AGP wavefunctions are calculated using the following Onishi formula \cite{onishi1966generator}.
\begin{eqnarray}
\braket{ F^{(\lambda)}|F^{(\mu)}}= \left.\exp \left(\frac{1}{2}\mathrm{tr}\left[ \ln (1+ F^{(\mu)}F^{(\lambda)T}x)\right] \right)\right|_{x^{n}} \label{OYovl}
\end{eqnarray}
Here, $x$ is an auxiliary variable, and $|_{x^n}$ denotes extraction of the coefficient of $x^n$ from the generating function. The matrix elements of the Hamiltonian and the gradients required for the variational optimization can be expressed in an analogous way using the Onishi formula.
Although AGP and LC-AGP can both be treated variationally using the Onishi formula, the practical limitations of LC-AGP are twofold. First, maintaining accuracy for larger systems or in strongly correlated regimes may require a rapidly increasing number of AGP components \cite{2015PhRvA..91f2504U}. Second, while the linear coefficients of an LC-AGP expansion are easily determined once the constituent AGPs are fixed, full variational optimization of the geminals entering each AGP is considerably more difficult, resembling a nonorthogonal multiconfigurational optimization and being prone to numerical difficulties.

Another representative approach to incorporate correlations between geminals is the antisymmetric product of geminals (APG) wavefunction \cite{kutzelnigg1964w, mcweeny1963density, kawasaki2025low}, which is constructed as a product of distinct geminals. For a system with $2n$ electrons, the APG wave function is given by
\begin{eqnarray}
\ket{\Psi_{\mathrm{APG}} } &=&  \hat{F}^{(1)}\cdots \hat{F}^{(n)} \ket{-} 
\end{eqnarray}
Since the APG wavefunction is constructed from a product of distinct geminals, the Onishi formula (Eq.~(\ref{OYovl})) cannot be applied straightforwardly. Therefore, we employ the classical Fischer formula \cite{fischer1994sums}, which expresses the product of $n$ commuting geminals as a signed linear combination of $n$-th powers of their linear combinations.
\begin{eqnarray}
\hat{F}^{(1)}\hat{F}^{(2)}\cdots\hat{F}^{(n)} = \frac{1}{2^{n-1}n!}\sum_{i_2=0}^1\cdots\sum_{i_n=0}^1 (-1)^{i_2 + \cdots + i_n}\left(\hat{F}^{(1)}+(-1)^{i_2}\hat{F}^{(2)}+\cdots + (-1)^{i_n}\hat{F}^{(n)}\right)^n  \label{fischer}
\end{eqnarray}

In tensor language, this is a Waring (symmetric-rank) decomposition of the corresponding polynomial.
This decomposition allows the APG wavefunction to be rewritten in LC-AGP form, making it possible to employ the Onishi formula (Eq.~(\ref{OYovl})).
APG has been reported to achieve higher accuracy compared to LC-AGP when
the same number of geminal types is employed \cite{kawasaki2025low}. However, as can be seen in Eq.~(\ref{fischer}), the number of LC-AGP terms generated by the Fischer decomposition grows exponentially with the number of electrons, leading to a prohibitive increase in computational cost.

This APG example illustrates the broader difficulty addressed in the present work. For general geminal-based wave functions, efficient deterministic evaluation is considerably less straightforward than for AGP or LC-AGP. While direct formulas are known in some cases, they are generally tractable only for special structured families, such as AGP, antisymmetrized product of strongly orthogonal geminals (APSG) \cite{PhysRevA.30.43, arai1960theorem, hurley1953molecular, Tokmachev:2016:PPG, tarumi2013accelerating, jeszenszki2015local, mcweeny1980molecular, PhysRevA.32.748, https://doi.org/10.1002/qua.560520225, surjan1999introduction, kapuy1960derivation, jeszenszki2014perspectives, nakatani2025construction}, or Richardson--Gaudin-type states \cite{richardson1963restricted, richardson1964exact, richardson1965exact, johnson2023single}; in more general settings, reducing the target state to an LC-AGP form remains one of the few broadly practical ways to exploit existing AGP overlap and Hamiltonian machinery efficiently.

Motivated by this viewpoint, in this work we introduce a class of antisymmetrized geminal power configuration interaction (AGP-CI) wavefunctions, formulated in a manner analogous to a CI expansion around an AGP reference. These wave functions incorporate inter-geminal correlations through one or more independent pair-creation operators while retaining a polynomial structure in commuting geminal operators. In this sense, AGP-CI is similar in spirit to APG, but it is designed so that its exact Waring decomposition into LC-AGP grows only polynomially with the number of electrons, rather than exponentially as in the APG case. Thus, AGP-CI provides a more expressive geminal ansatz than AGP while remaining much more manageable than a direct APG-to-LC-AGP expansion.

We then take a further step and exploit border-rank approximations of the AGP-CI polynomial. Introducing a small deformation parameter $\tau$, we reorganize the AGP-CI wave function into an even more compact LC-AGP representation, which we denote AGP-CI$\tau$. The role of the border-rank construction is therefore practical rather than purely formal. On the one hand, AGP-CI is the physically motivated target ansatz, whereas LC-AGP is the computational representation in which overlaps and observables can be evaluated efficiently; the border-rank approximation therefore provides a compressed bridge between the two. On the other hand, by reducing the number of AGP components that must be handled explicitly, it offers a more compact variational form for directly optimizing AGP-CI$\tau$ and can also reduce the cost of subsequent correlation treatments based on the optimized AGP-CI$\tau$ state. In addition, because the border-rank representation embeds AGP-CI into the LC-AGP manifold in a structured way, it can also serve as a physically informed initial guess for subsequent LC-AGP optimization, allowing one to explore the region of LC-AGP space near the optimized AGP-CI state.

CI-like extensions of AGP have been considered in earlier work. Henderson \textit{et~al.}~formulated the AGP-based CI wavefunction by applying annihilation operators to an AGP state and demonstrated its effectiveness for the pairing Hamiltonian \cite{henderson2019geminal}. Dutta \textit{et~al.}~proposed an LC-AGP framework in which the original AGP is replaced by AGP states with modified geminal coefficients, followed by a linear combination of these states \cite{dutta2021construction}. The AGP-CI wavefunctions studied in this work are conceptually aligned with these approaches, as they likewise involve replacing part of the AGP structure with different operators. The key contribution of the present work is the introduction of a border-rank approximation applied to AGP-CI, yielding the AGP-CI$\tau$ ansatz. This construction enables a much more compact representation and leads to a substantial reduction in computational cost.

The rest of this article is organized as follows. Section \ref{s2} introduces the AGP-CI framework together with the border-rank decompositions of the AGP-CI polynomial, and presents the formulation of the compact AGP-CI, referred to as AGP-CI$\tau$. Section \ref{s3} reports the results obtained by applying the proposed AGP-CI$\tau$ wavefunction to the Hubbard model and to small molecules. Finally, Section \ref{s4} provides concluding remarks.

\section{Theory} \label{s2}

To go beyond AGP and incorporate correlations between geminals, we consider a CI-like expansion in which one or more geminals in the AGP reference are replaced by independent pair-creation operators. Truncating the expansion at the level of $k$ pair replacements defines the AGP-CI($k$) family; in particular, AGP-CIS, AGP-CID, and AGP-CIT retain up to one-, two-, and three-pair replacements.
\begin{eqnarray}
\ket{\Psi_{\mathrm{AGP-CIS}} } &=& \hat{F}^n\ket{-}  + \hat{C}^{(1)}\hat{F}^{n-1}  \ket{-}  \\
\ket{\Psi_{\mathrm{AGP-CID}} } &=& \hat{F}^n\ket{-}  + \hat{C}^{(1)}\hat{C}^{(2)}\hat{F}^{n-2} \ket{-}  \\
\ket{\Psi_{\mathrm{AGP-CIT}} } &=& \hat{F}^n\ket{-}  + \hat{C}^{(1)}\hat{C}^{(2)}\hat{C}^{(3)}\hat{F}^{n-3} \ket{-}  
\end{eqnarray}
Here, each $\hat C^{(i)}$ is a pair-creation operator parameterized by an antisymmetric matrix $C^{(i)}$,
\begin{equation}
\hat C^{(i)} \equiv \sum_{ab}^{2m} C^{(i)}_{ab}\,
\hat c_a^\dagger \hat c_b^\dagger,
\qquad C^{(i)}_{ab}=-C^{(i)}_{ba}.
\end{equation}
%
%
Although double and triple excitations are conventionally represented by fourth- and sixth-order tensors, respectively, we express them as products of geminals parameterized by antisymmetric matrices for computational efficiency. In this form of the wavefunctions, for example, the "singles
+ doubles" truncation can be absorbed into a double-replacement ansatz, i.e AGP-CISD becomes equivalent to AGP-CID. See Appendix \ref{app1} for details.

Similar to APG, the AGP-CI wavefunctions cannot be evaluated directly using Eq.~(\ref{OYovl}). We therefore first rewrite AGP-CI in the LC-AGP form.
From a polynomial point of view, an AGP-CI wavefunction may be regarded as a homogeneous polynomial of degree $n$ in the commuting geminal operators $\hat{F}, \hat{C}^{(1)}, \hat{C}^{(2)}, \dots$. An LC-AGP of the form
\[
\sum_{r=1}^{R} \bigl( \hat{F}^{(r)} \bigr)^n
\]
is then a sum of the $R$ pure $n$-th powers of linear combinations of these geminals. Expressing a given AGP-CI state in LC-AGP form is therefore equivalent to finding a rank-$R$ decomposition of the associated homogeneous polynomial, this is precisely a Waring decomposition of a homogeneous polynomial; the minimal
such $R$ is the Waring rank (equivalently, the symmetric tensor rank)
\cite{Landsberg2012}.

Specializing this point of view to the AGP-CI building blocks, we obtain
explicit rank decompositions for monomials 
$\hat{C}^{(1)}\hat{F}^{n-1}$,
$\hat{C}^{(1)}\hat{C}^{(2)}\hat{F}^{n-2}$, and
$\hat{C}^{(1)}\hat{C}^{(2)}\hat{C}^{(3)}\hat{F}^{n-3}$.
By applying the Fischer formula to monomials with repeated variables and
collecting terms, one finds that

\begin{eqnarray}
\hat{C}^{(1)}\hat{F}^{n-1} &=& \sum_{k=0}^{n-1} \frac{1}{n} \frac{1}{(n-1-k)!k!}\frac{(-1)^{k}}{2^{n-1}} \left( \hat{C}^{(1)} + (n-1-2k)\hat{F} \right)^n \label{poly1} \\ 
\hat{C}^{(1)}\hat{C}^{(2)}\hat{F}^{n-2} &=& \sum_{k=0}^{n-2} \sum_{l=0}^1 \frac{1}{n(n-1)} \frac{1}{(n-2-k)!k!}\frac{(-1)^{k+l}}{2^{n-1}} \left( \hat{C}^{(1)} + (-1)^l\hat{C}^{(2)} + (n-2-2k)\hat{F} \right)^n \nonumber \\
\label{poly2} \\ 
\hat{C}^{(1)}\hat{C}^{(2)}\hat{C}^{(3)}\hat{F}^{n-3} &=& \sum_{k=0}^{n-3} \sum_{l=0}^1 \sum_{m=0}^1\frac{1}{n(n-1)(n-2)} \frac{1}{(n-3-k)!k!}\frac{(-1)^{k+l+m}}{2^{n-1}} \nonumber \\
&&\times  \left( \hat{C}^{(1)} + (-1)^l\hat{C}^{(2)} + (-1)^m\hat{C}^{(3)} + (n-3-2k)\hat{F} \right)^n \label{poly3}
\end{eqnarray}
Each term on the right-hand side is a pure $n$-th power of a linear
combination of geminals, i.e., an AGP-type factor. Hence, Eqs.~(\ref{poly1})--(\ref{poly3})
provide LC-AGP representations with $n$, $2(n-1)$, and $4(n-2)$ AGP terms,
respectively.
In the polynomial language, this means that monomials 
$\hat{C}^{(1)}\hat{F}^{n-1}$,
$\hat{C}^{(1)}\hat{C}^{(2)}\hat{F}^{n-2}$, and
$\hat{C}^{(1)}\hat{C}^{(2)}\hat{C}^{(3)}\hat{F}^{n-3}$
have Waring rank at most $n$, $2(n-1)$, and $4(n-2)$. 
Moreover, these bounds are tight. Under the identification $\hat C^{(i)}\mapsto x_i$ and $\hat F\mapsto x_{k+1}$, the above monomials correspond to
$x_1 x_2^{\,n-1}$,
$x_1 x_2 x_3^{\,n-2}$, and
$x_1 x_2 x_3 x_4^{\,n-3}$,
The general formula for the Waring rank of a monomial
$x_1^{a_1}\cdots x_m^{a_m}$ with $1 \le a_1 \le \cdots \le a_m$ is
$\prod_{i=2}^m (a_i+1)$
, which yields the lower bounds
$n$, $2(n-1)$, and $4(n-2)$, respectively\cite{carlini2012solution}. Therefore the ranks above are optimal.

It is often sufficient to approximate a polynomial rather than represent it exactly. This motivates the \emph{border rank} \cite{landsberg2010ranks}, which captures the minimal rank achievable in a limiting (approximate) sense.
In polynomial decomposition, the \emph{border rank} of a homogeneous polynomial $f$ is the smallest integer $R$ such that $f$ can be obtained as a limit of polynomials of Waring rank at most $R$. A simple example illustrating its difference with rank is the monomial $x_1 x_2^{\,n-1}$: its Waring rank is $n$, yet its border rank is $2$, since
\begin{equation}
x_1 x_2^{\,n-1}
= \frac{1}{n}\lim_{t\to 0}\frac{(x_2+t x_1)^n - x_2^n}{t},
\end{equation}
i.e., it arises as a limit of differences of two $n$-th powers. Thus border rank can be much smaller than the Waring rank.

While Eqs.~(\ref{poly1})--(\ref{poly3}) provide exact Waring decompositions, the same monomials can be obtained as limits of only a few $n$-th powers.  This yields the following border-rank forms in the limit $\tau\to 0$:
\begin{eqnarray}
\hat{C}^{(1)}\hat{F}^{n-1} &=&  \lim_{\tau \to 0} \frac{1}{2n\tau} \left[ \left(\hat{F} + \tau \hat{C}^{(1)} \right)^n - \left(\hat{F} - \tau \hat{C}^{(1)}  \right)^n  \right]  \label{polyt1}\\
 \hat{C}^{(1)}\hat{C}^{(2)}\hat{F}^{n-2} &=&  \lim_{\tau \to 0} \frac{1}{4n(n-1)\tau^2} \left[ \left(\hat{F} + \tau( \hat{C}^{(1)} + \hat{C}^{(2)} ) \right)^n +\left(\hat{F} - \tau( \hat{C}^{(1)} + \hat{C}^{(2)} ) \right)^n  \right.  \nonumber \\
&& \hspace{2.65cm} \left. - \left(\hat{F} + \tau( \hat{C}^{(1)} - \hat{C}^{(2)} ) \right)^n- \left(\hat{F} - \tau( \hat{C}^{(1)} - \hat{C}^{(2)} ) \right)^n \right] \label{polyt2} \\
\hat{C}^{(1)}\hat{C}^{(2)}\hat{C}^{(3)}\hat{F}^{n-3} &=& \lim_{\tau \to 0} \frac{1}{8n(n-1)(n-2)\tau^3} \left[ \left(\hat{F} + \tau( \hat{C}^{(1)} + \hat{C}^{(2)} + \hat{C}^{(3)}) \right)^n  +\left(\hat{F} - \tau( - \hat{C}^{(1)} + \hat{C}^{(2)} + \hat{C}^{(3)}  ) \right)^n  \right.  \nonumber \\
&&  \hspace{3.8cm}  + \left(\hat{F} - \tau( \hat{C}^{(1)} - \hat{C}^{(2)} + \hat{C}^{(3)}) \right)^n +\left(\hat{F} - \tau( \hat{C}^{(1)} + \hat{C}^{(2)} - \hat{C}^{(3)}  ) \right)^n \nonumber \\
&&  \hspace{3.8cm}  - \left(\hat{F} - \tau(  \hat{C}^{(1)} + \hat{C}^{(2)} + \hat{C}^{(3)}  ) \right)^n - \left(\hat{F} + \tau( -\hat{C}^{(1)} + \hat{C}^{(2)} + \hat{C}^{(3)}) \right)^n  \nonumber \\
&&  \hspace{3.8cm}  \left. - \left(\hat{F} + \tau(  \hat{C}^{(1)} - \hat{C}^{(2)} + \hat{C}^{(3)}  ) \right)^n - \left(\hat{F} + \tau( \hat{C}^{(1)} + \hat{C}^{(2)} - \hat{C}^{(3)}) \right)^n   \right] \nonumber \\ \label{polyt3}
\end{eqnarray}
Eqs.~(\ref{polyt1})--(\ref{polyt3}) express the AGP-CI monomials as limits of finite-difference combinations of pure $n$-th powers. In particular, Eq.~(\ref{polyt1}) involves 2 AGP factors, Eq.~(\ref{polyt2}) involves 4, and Eq.~(\ref{polyt3}) involves 8.  In contrast, the exact Waring decompositions in Eqs.~(\ref{poly1})--(\ref{poly3}) require $n$, $2(n-1)$, and $4(n-2)$ AGP terms, respectively.  Thus, the border-rank forms reduce the number of AGP terms from $\mathcal{O}(n)$ to $\mathcal{O}(1)$, which is expected to lower the cost of evaluating overlaps and Hamiltonian matrix elements. In practical numerical calculations, rather than taking the limit $\tau \to 0$, we employ a small but finite $\tau$ so that Eqs.~(\ref{polyt1})--(\ref{polyt3}) provide controlled finite-$\tau$ approximations that recover the target monomials in the $\tau\to 0$ limit. We denote the resulting forms as the AGP-CI$\tau$ wavefunctions. For small $\tau$, these finite-difference expressions can be numerically ill-conditioned due to subtraction of nearly equal AGP factors. To mitigate this issue, we remove the AGP term from the AGP-CI$\tau$ wavefunctions as follows.
\begin{eqnarray}
\ket{\Psi_{\mathrm{AGP-CIS}\tau} } &=&   \frac{1}{2n\tau} \left[ \left(\hat{F} + \tau \hat{C}^{(1)} \right)^n - \left(\hat{F} - \tau \hat{C}^{(1)}  \right)^n  \right] \ket{-} \label{agpcist} \\
\ket{\Psi_{\mathrm{AGP-CID}\tau} } &=&   \frac{1}{4n(n-1)\tau^2} \left[ \left(\hat{F} + \tau( \hat{C}^{(1)} + \hat{C}^{(2)} ) \right)^n +\left(\hat{F} - \tau( \hat{C}^{(1)} + \hat{C}^{(2)} ) \right)^n  \right.  \nonumber \\
&& \hspace{1.9cm} \left. - \left(\hat{F} + \tau( \hat{C}^{(1)} - \hat{C}^{(2)} ) \right)^n- \left(\hat{F} - \tau( \hat{C}^{(1)} - \hat{C}^{(2)} ) \right)^n \right]  \ket{-}  \label{agpcidt} \\
\ket{\Psi_{\mathrm{AGP-CIT}\tau} } &=& \frac{1}{8n(n-1)(n-2)\tau^3} \left[ \left(\hat{F} + \tau( \hat{C}^{(1)} + \hat{C}^{(2)} + \hat{C}^{(3)}) \right)^n  +\left(\hat{F} - \tau( - \hat{C}^{(1)} + \hat{C}^{(2)} + \hat{C}^{(3)}  ) \right)^n  \right.  \nonumber \\
&&  \hspace{3.1cm}  + \left(\hat{F} - \tau( \hat{C}^{(1)} - \hat{C}^{(2)} + \hat{C}^{(3)}) \right)^n +\left(\hat{F} - \tau( \hat{C}^{(1)} + \hat{C}^{(2)} - \hat{C}^{(3)}  ) \right)^n \nonumber \\
&&  \hspace{3.1cm}  - \left(\hat{F} - \tau(  \hat{C}^{(1)} + \hat{C}^{(2)} + \hat{C}^{(3)}  ) \right)^n - \left(\hat{F} + \tau( -\hat{C}^{(1)} + \hat{C}^{(2)} + \hat{C}^{(3)}) \right)^n  \nonumber \\
&&  \hspace{3.1cm}  \left. - \left(\hat{F} + \tau(  \hat{C}^{(1)} - \hat{C}^{(2)} + \hat{C}^{(3)}  ) \right)^n - \left(\hat{F} + \tau( \hat{C}^{(1)} + \hat{C}^{(2)} - \hat{C}^{(3)}) \right)^n   \right] \ket{-}   \nonumber \\ \label{agpcitt}
\end{eqnarray}
Table \ref{agp_term} shows the number of AGP terms and the number of variational parameters for each wavefunction. The computational cost of AGP in this work scales as $\mathcal{O}(m^5)$. For wavefunctions composed of $k$ AGP terms, the cost becomes $\mathcal{O}(k^2 m^5)$. By applying the border-rank approximation to AGP-CI, the number of AGP terms in AGP-CI$\tau$ is reduced from $\mathcal{O}(n)$ to $\mathcal{O}(1)$, leading to a substantial reduction in computational cost, corresponding to an approximate $n^2$ speedup.
\begin{table}
\caption{Number of AGP terms and variational parameters.}
{\begin{tabular}{l  ccc} \hline
    & AGP term & variational parameters    \\ \hline
AGP & 1 & $m(2m-1) $   \\  \hline
AGP-CIS$\tau$  & 2 & $2m(2m-1)$  \\
AGP-CID$\tau$    & 4 & $3m(2m-1)$   \\ 
AGP-CIT$\tau$    & 8  & $ 4m(2m-1)$ \\
LC-AGP ($K=k$)      & $k$ & $km(2m-1)$   \\ \hline
\end{tabular}}
\label{agp_term}
\end{table}

\section{Demonstrative applications} \label{s3}

In this section, we present the results of applying the AGP-CI$\tau$ wavefunctions, along with other geminal-based wavefunctions, to the one-dimensional Hubbard model with periodic boundary conditions and to the small molecules H$_2$O and N$_2$.
The Hamiltonian of the one-dimensional Hubbard model with periodic boundary conditions is given by
\begin{eqnarray}
\hat{H} = -t \sum_{i=1}^{m} \sum_{\sigma=\uparrow,\downarrow} 
\left( \hat{c}^\dagger_{i,\sigma} \hat{c}_{i+1,\sigma} + \hat{c}^\dagger_{i+1,\sigma} \hat{c}_{i,\sigma} \right)
+ U \sum_{i=1}^{m} \hat{n}_{i,\uparrow} \hat{n}_{i,\downarrow}
\end{eqnarray}
where $t$ and $U$ denote the hopping parameter and on-site Coulomb repulsion, respectively, $c^\dagger_{i,\sigma}$ ($c_{i,\sigma}$) is the creation (annihilation) operator for an electron with spin $\sigma$ at site $i$, and $n_{i,\sigma} = c^\dagger_{i,\sigma} c_{i,\sigma}$ is the number operator.
Calculations for H$_2$O and N$_2$ were performed using the STO-6G basis set. The integrals of the molecular Hamiltonian were computed using PySCF, and the results were compared with the exact diagonalization results obtained with the H$\Phi$ package \cite{kawamura2017quantum}. The optimization was performed using the conjugate gradient method implemented in SciPy. The variational parameters were initialized randomly, and each calculation was repeated approximately 10 times to avoid being trapped in local minima.

First, we applied the AGP-CID wavefunction to the Hubbard model at half filling with $U/t = 10$. The results for the total energies, compared with those from APG, LC-AGP, and AGP, are summarized in Table \ref{hubbard}. For the 12-electron system, the accuracy of LC-AGP is observed to decrease significantly. Conversely, AGP-CID maintains accuracy even for the 12-electron system. These results indicate that AGP-CI type wavefunctions can overcome the limitation of LC-AGP, where the number of AGP terms needed to achieve a given accuracy grows with the system size. We emphasize that this limitation of LC-AGP is likely due to optimization challenges rather than a fundamental limitation of the ansatz itself.
\begin{table}
\caption{Total energy and the energy error relative to the exact solution for the Hubbard model ($U/t=10$, half filling, $E/t$).}
{\begin{tabular}{l  cccccc} \hline
 Number of electrons & 8 &  & 10 & & 12 & \\ \hline
 Exact & -2.176688 &  $\Delta$  & -2.703691 & $\Delta$ & -3.232383 & $\Delta$ \\  \hline
 APG   & -1.917334 & 0.259354 & -2.376477 & 0.327214 & -2.842129 & 0.390254 \\
 LC-AGP ($K=n$) & -1.888135 & 0.288553 & -2.317048 & 0.386643 & -2.480521 & 0.751862 \\
 AGP            & -1.626127 & 0.550561 & -1.994161 & 0.709530 & -2.388118 & 0.844265 \\ \hline
 AGP-CID        & -1.838595 & 0.338093 & -2.229593 & 0.474098 & -2.613798 & 0.618585 \\ \hline
\end{tabular}}
\label{hubbard}
\end{table}

\begin{figure}
\centering
{\resizebox*{10cm}{!}{\includegraphics{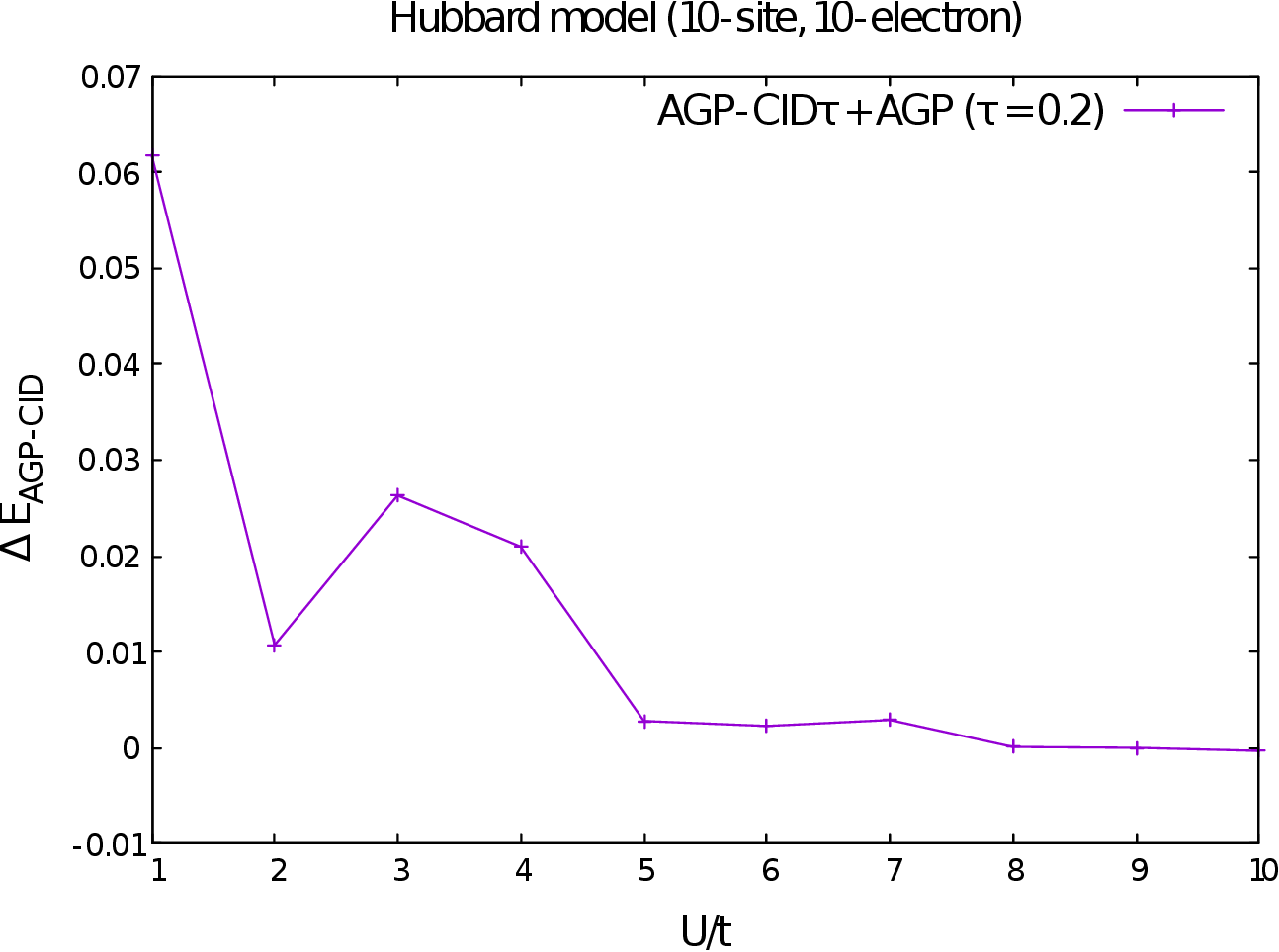}}}
\caption{Total energy error with respect to AGP-CID for AGP-CID$\tau$ + AGP with $\tau = 0.2$ in the Hubbard model (10-site, 10-electron system, $\Delta E_{\mathrm{AGP-CID}}=E_{\mathrm{AGP-CID}\tau+\mathrm{AGP}}-E_\mathrm{AGP-CID}$, $E/t$).} 
\label{t_0.2}
\end{figure}

Next, using the results obtained in Table \ref{hubbard}, we investigate the appropriate value of $\tau$ for the AGP-CI$\tau$ wavefunctions. Table \ref{suitable_t} shows the results of total energy calculations using the AGP-CID$\tau$ + AGP wavefunctions, where the parameter $\tau$ is varied from 0.1 to 1. In these calculations, we employed the optimized parameters $F$, $C^{(1)}$, and $C^{(2)}$ obtained from the AGP-CID results in Table \ref{hubbard}. We note that AGP-CID$\tau$ + AGP, rather than AGP-CID$\tau$, was used so that the optimized AGP-CID parameters could be directly utilized. The explicit form of the AGP-CID$\tau$+AGP wavefunctions is provided in Appendix \ref{app2}. Appendix \ref{app2} also presents a comparison between AGP-CI$\tau$ and AGP-CI$\tau$+AGP results. From results in Table \ref{suitable_t}, it can be seen that $\tau=0.2$ yields the lowest total energy for the 8-electron and 10-electron systems, and for the 12-electron system, it gives the second-lowest total energy after $\tau=0.1$. Therefore, we set $\tau=0.2$ in all AGP-CI$\tau$ calculations presented in this paper. Although the optimal value of $\tau$ is expected to depend on the system size, examining the denominators in Eqs.~(\ref{agpcist})–(\ref{agpcitt}) suggests that choosing $\tau$ on the order of $1/n$ avoids division by excessively small numbers and is therefore unlikely to cause numerical instability. In the present work, since we consider systems with at most 20 electrons, fixing $\tau = 0.2$ is not expected to significantly degrade accuracy across this range of system sizes. For the 10-site, 10-electron system ($n=5$, $1/n=0.2$) in the Hubbard model, we vary $U/t$ from 1 to 10 and evaluate the error relative to AGP-CID using AGP-CID$\tau$ + AGP with $\tau=0.2$. Here as well, for each value of $U/t$, the parameters obtained from the optimized AGP-CID calculation are used in AGP-CID$\tau$ + AGP. The results are shown in Figure \ref{t_0.2}. For all values of $U/t$, the error relative to AGP-CID remains moderate, and it becomes smaller as $U/t$ increases. Although some non-monotonic variation is observed in the error at small $U/t$, the overall magnitude of the error remains small. Since the present work is aimed at developing methods for strongly correlated systems, the choice $\tau = 0.2$ appears to be appropriate. Results for the case where $\tau$ is treated as a variational parameter are shown in Appendix \ref{app3}.
\begin{table}
\caption{Total energy and the energy error relative to the exact solution for the Hubbard model ($U/t=10$, half filling, $E/t$)}
{\begin{tabular}{l  ccccccc} \hline
Number of electrons   & 8 &   & 10 & & 12 &  \\ \hline
AGP-CID & -1.838595 & $\Delta$  & -2.229593 & $\Delta$ & -2.613798 & $\Delta$ \\  \hline
$\tau$ (AGP-CID$\tau$ + AGP) & & & & & &  \\  \hline
0.1           & -1.838612 & -0.000017  & -2.229765  & -0.000171 & -2.613602 & 0.000196 \\
0.2           & -1.838636 & -0.000041  & -2.229893  & -0.000300 & -2.611445 & 0.002353  \\
0.3           & -1.838580 & 0.000014  &  -2.228817  &  0.000775 & -2.602608 & 0.011190 \\
0.4           & -1.838304 & 0.000291  &  -2.224602 &  0.004990  & -2.579214 & 0.034584 \\
0.5           & -1.837606 & 0.000989  &  -2.214547 &  0.015046  & -2.530282 & 0.083516 \\
0.6           & -1.836231 & 0.002363  &  -2.195202 &  0.034391  & -2.441941 & 0.171856 \\
0.7           & -1.833867 & 0.004728  &  -2.162410 &  0.067183  & -2.297933 & 0.315865 \\
0.8           & -1.830142 & 0.008452  &  -2.111383 &  0.118210  & -2.080550 & 0.533248 \\
0.9           & -1.824633 & 0.013962  &  -2.036824 & 0.192769   & -1.772114 & 0.841684 \\
1.0           & -1.816857 & 0.021737  &  -1.933109 & 0.296484   & -1.357003 & 1.256795 \\ \hline
\end{tabular}}
\label{suitable_t}
\end{table}

\begin{figure}
\centering
{\resizebox*{10cm}{!}{\includegraphics{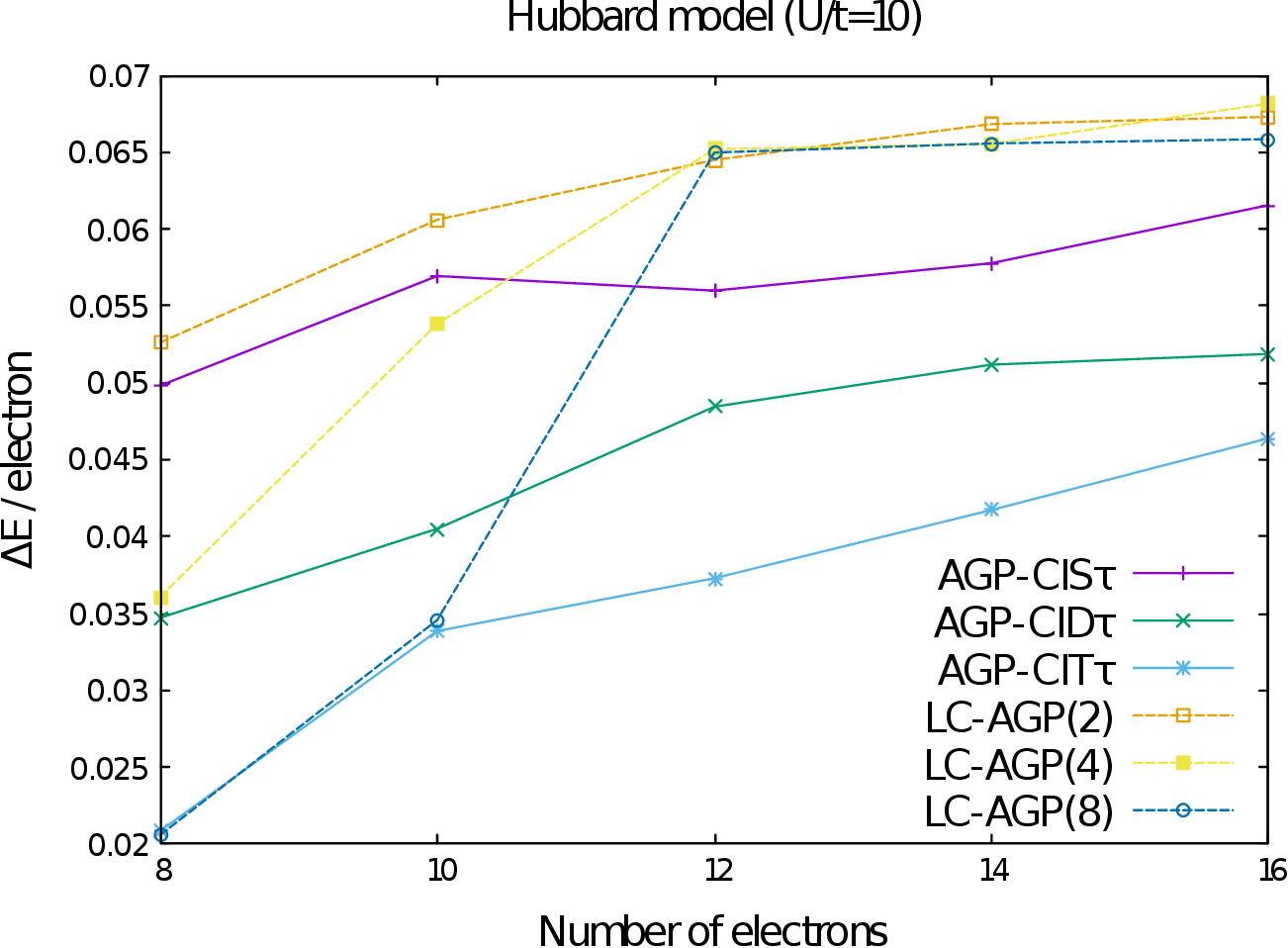}}}
\caption{Total energy error per electron for the Hubbard model ($U/t = 10$, half filling, $E/t$).} 
\label{energy_u10}
\end{figure}

Table \ref{hubbard_10} summarizes the total energy results obtained with AGP-CI$\tau$ and LC-AGP for the half-filled Hubbard model at $U/t = 10$. The results for $U/t = 1$ and $-1$ are provided in the supplementary material, where a similar trend is observed. For LC-AGP, calculations were performed with $K=2$, 4, and 8 to match the number of AGP terms in AGP-CIS$\tau$, AGP-CID$\tau$, and AGP-CIT$\tau$, respectively. In addition, Figure \ref{energy_u10} shows the errors per electron relative to the exact solution. It can be seen that the results of LC-AGP deteriorate for systems with 12 or more electrons. For systems with 12 or more electrons, there is almost no difference in accuracy among LC-AGP calculations with $K=2$, 4, and 8, suggesting that the number of AGP terms required to maintain high accuracy increases rapidly. On the other hand, the AGP-CI$\tau$ wavefunctions maintain high accuracy even as the number of electrons increases, and it is clearly seen that the accuracy improves progressively as the CI expansion is extended from AGP-CIS$\tau$ to AGP-CID$\tau$ and AGP-CIT$\tau$. Although LC-AGP and AGP-CI employ the same number of AGP terms, the variational space of LC-AGP is larger. Therefore, if the optimization were sufficiently converged, LC-AGP should yield results comparable to or better than those of AGP-CI. The present results thus reflect the difficulty of optimizing LC-AGP.
\begin{table}
\caption{Total energy for the Hubbard model ($U/t=10$, half filling, $E/t$).}
{\begin{tabular}{l  ccccc} \hline
Number of electrons    & 8 & 10 & 12 & 14 & 16   \\ \hline
Exact & -2.176688 & -2.703691 & -3.232383 & -3.762854 & -4.294284   \\  \hline
AGP-CIS$\tau$   & -1.778255 & -2.134351 & -2.560479 & -2.954145 & -3.309395 \\
AGP-CID$\tau$   & -1.899057 & -2.298660 & -2.651037 & -3.046418 & -3.464406 \\ 
AGP-CIT$\tau$   & -2.009888 & -2.365190 & -2.784994 & -3.179332 & -3.552057 \\
LC-AGP ($K=2$) & -1.755789 & -2.097645 & -2.458466 & -2.827311 & -3.217516 \\
LC-AGP ($K=4$) & -1.888135 & -2.165343 & -2.450136 & -2.845319 & -3.203955 \\
LC-AGP ($K=8$) & -2.012119 & -2.358130 & -2.452696 & -2.844895 & -3.240758 \\ \hline
\end{tabular}}
\label{hubbard_10}
\end{table}

\begin{figure}
\centering
{\resizebox*{10cm}{!}{\includegraphics{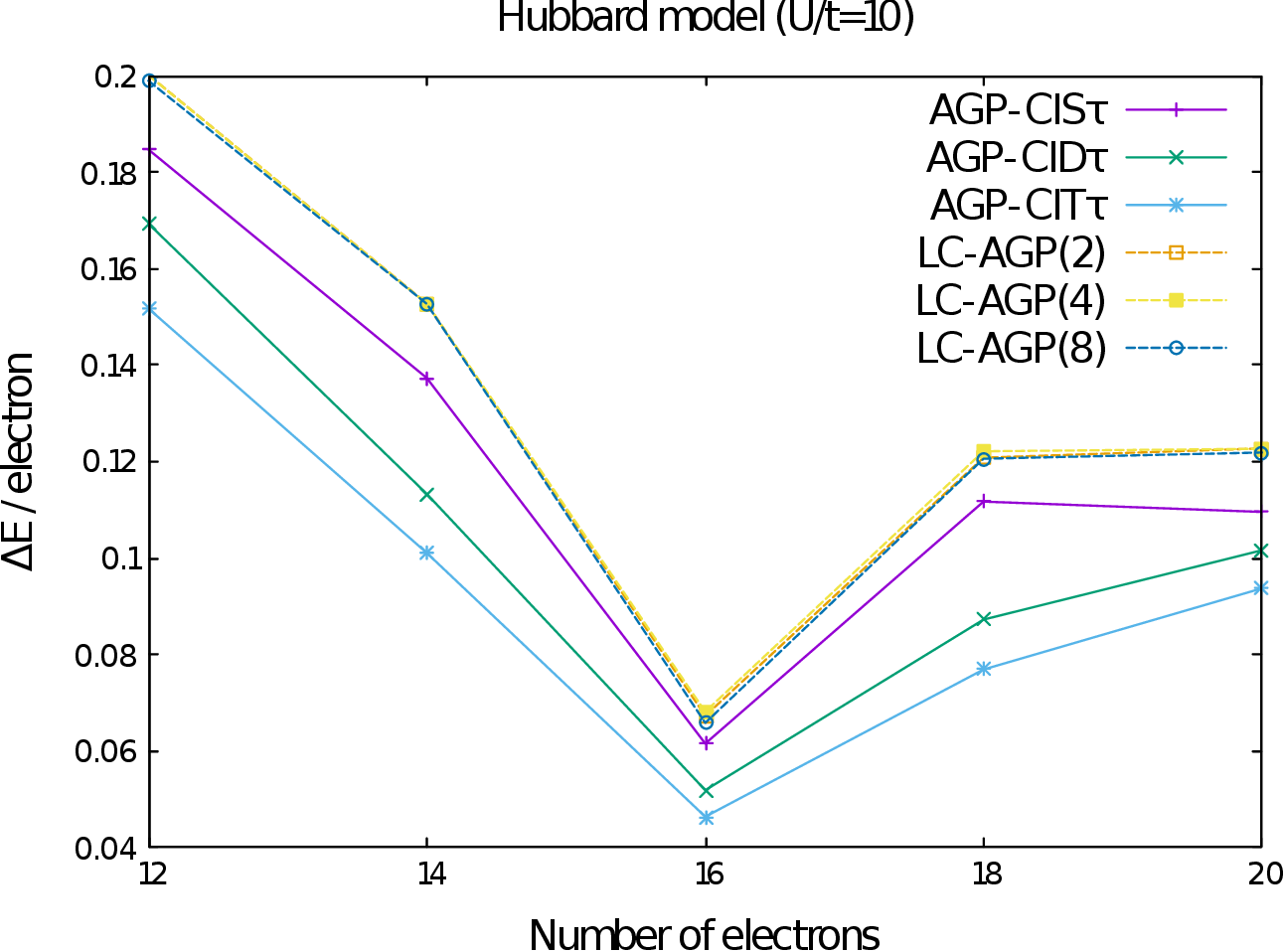}}}
\caption{Total energy error per electron for the Hubbard model ($U/t = 10$, 16-site system, $E/t$).} 
\label{energy_nh}
\end{figure}

Table \ref{hubbard_nh} shows the total energy results for the 16-site Hubbard model at $U/t = 10$ obtained by varying the electron number, thereby exploring fillings both at and away from half filling. Figure \ref{energy_nh} shows the errors per electron relative to the exact solution.
Beyond half filling (16-electron, 16-site system), the AGP-CI$\tau$ wavefunctions also yield better results than LC-AGP.
In LC-AGP, increasing the number of AGP terms leads to little change in accuracy, whereas in AGP-CI$\tau$ a clear and systematic improvement is observed as the excitation level is increased from S to D and T.

\begin{table}
\caption{Total energy for the Hubbard model ($U/t=10$, 16-site system, $E/t$).}
{\begin{tabular}{l  ccccc} \hline
Number of electrons    & 12 & 14 & 16 & 18 & 20   \\ \hline
Exact & -9.983329 & -7.620396 & -4.294284 & 12.379604 & 30.016671   \\  \hline
AGP-CIS$\tau$   & -7.766993 & -5.698119 & -3.309395 & 14.389856 & 32.207303 \\
AGP-CID$\tau$   & -7.951420 & -6.034239 & -3.464406 & 13.950029 & 32.049225 \\ 
AGP-CIT$\tau$   & -8.162363 & -6.205591 & -3.552057 & 13.763468 & 31.890719 \\
LC-AGP ($K=2$) & -7.581166 & -5.481826 & -3.217516 & 14.553834 & 32.471286 \\
LC-AGP ($K=4$) & -7.581814 & -5.482171 & -3.203955 & 14.577661 & 32.469072 \\
LC-AGP ($K=8$) & -7.596452 & -5.482895 & -3.240758 & 14.549856 & 32.454390 \\ \hline
\end{tabular}}
\label{hubbard_nh}
\end{table}

We further extended our validation to small molecular systems, specifically H$_2$O and N$_2$. The molecular coordinates used here, which correspond to approximately equilibrium nuclear positions, are H$_2$O [O(0, 0, 0), H($-1.809$, 0, 0), H(0.453549, 1.751221, 0)] and N$_2$ [N(0, 0, 0), N(0, 0, 2.3)], with all values given in bohr.

Table~\ref{molecule} shows the total energy results of applying AGP-CI$\tau$ and LC-AGP to H$_2$O and N$_2$. In the case of H$_2$O, LC-AGP is more accurate, while for N$_2$, AGP-CI$\tau$ provides superior accuracy. This behavior is consistent with the results for the Hubbard model, where the accuracy of LC-AGP decreases as the number of electrons increases and as electron correlation becomes stronger. AGP-CI$\tau$ is found to provide good accuracy in molecular systems as well.
\begin{table}
\caption{Total energy and the energy error relative to the exact solution for the H$_2$O and N$_2$ (STO-6G, equilibrium bond length, energy in Hartree).}
{\begin{tabular}{l  cccc} \hline
    & H$_2$O & $\Delta$ E (H$_2$O) & N$_2$ & $\Delta$ E (N$_2$)   \\ \hline
Exact & -75.72871 & - & -108.7265 & -    \\  \hline
AGP-CIS$\tau$   & -75.72250 & 0.006202 & -108.6463 & 0.080211  \\
AGP-CID$\tau$   & -75.72694 & 0.001769 & -108.6891 & 0.037437 \\ 
AGP-CIT$\tau$   & -75.72794 & 0.000765 & -108.7033 & 0.023187 \\
LC-AGP ($K=2$) & -75.71970 & 0.009007 & -108.6173 & 0.109159 \\
LC-AGP ($K=4$) & -75.72708 & 0.001624 & -108.6735 & 0.052971 \\
LC-AGP ($K=8$) & -75.72861 & 0.000098 & -108.6771 & 0.049380 \\ \hline
\end{tabular}}
\label{molecule}
\end{table}

\begin{figure}
\centering
{\resizebox*{10cm}{!}{\includegraphics{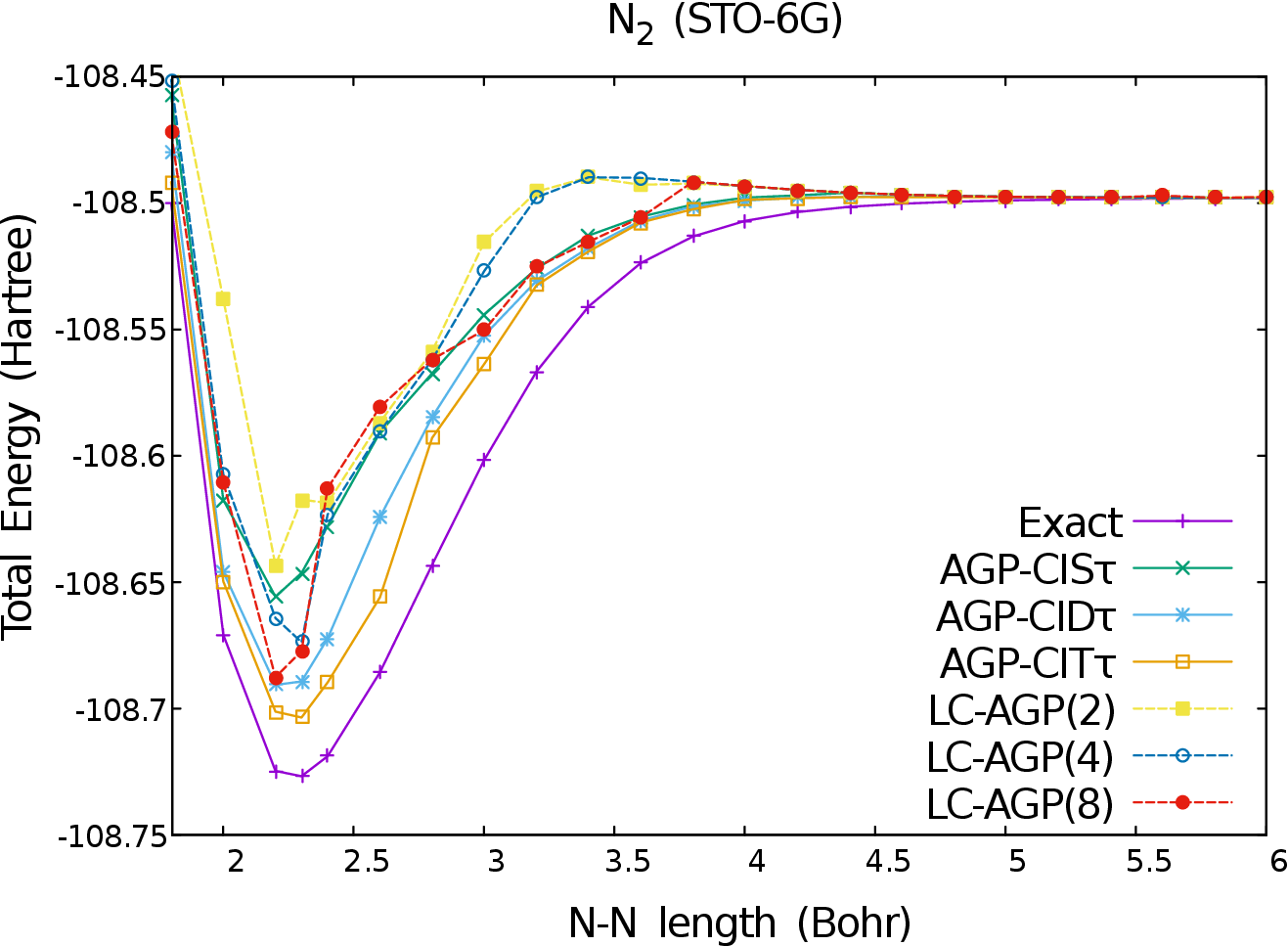}}}
\caption{Potential energy curve of N$_2$ (STO-6G).} 
\label{pe_n2}
\end{figure}

For N$_2$, we also examined bond lengths away from equilibrium, and the resulting potential energy curve is shown in Figure \ref{pe_n2}. For AGP-CI$\tau$, the accuracy systematically improves as one moves from AGP-CIS$\tau$ to AGP-CID$\tau$ and AGP-CIT$\tau$, resulting in a smooth and well-behaved potential energy curve over the entire range of bond distances. In contrast, LC-AGP exhibits noticeable irregularities, particularly in the vicinity of the equilibrium bond length, reflecting difficulties in the underlying optimization. These irregularities make it challenging to obtain a smooth and well-defined potential energy curve. This comparison demonstrates a practical advantage of AGP-CI$\tau$ beyond energetic accuracy, which stems from its numerical robustness.

\begin{figure}
\centering
{\resizebox*{10cm}{!}{\includegraphics{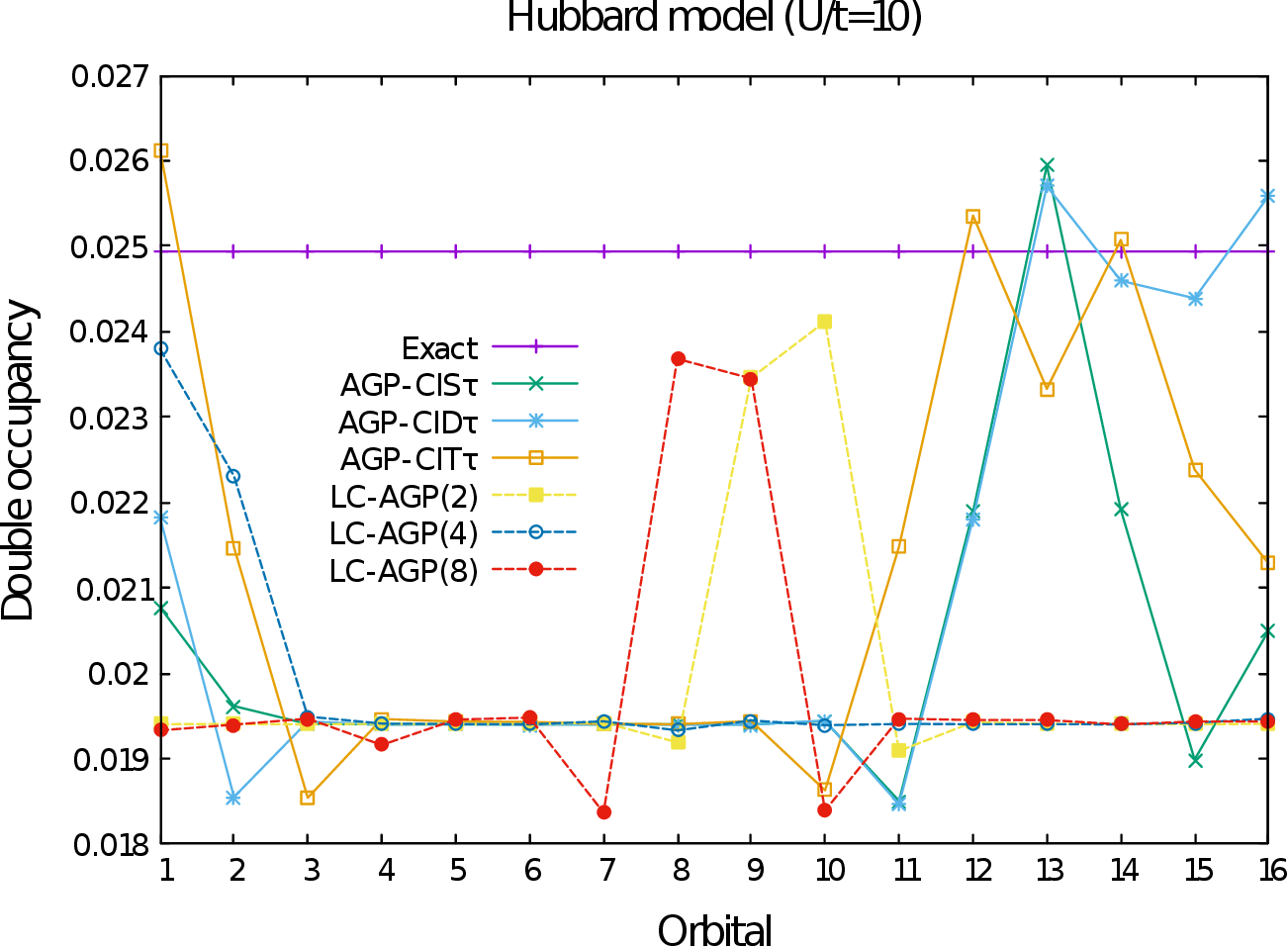}}}
\caption{Double occupancy for the Hubbard model ($U/t = 10$, 16-site, 16-electron system).} 
\label{do_hubbard}
\end{figure}

\begin{figure}
\centering
{\resizebox*{10cm}{!}{\includegraphics{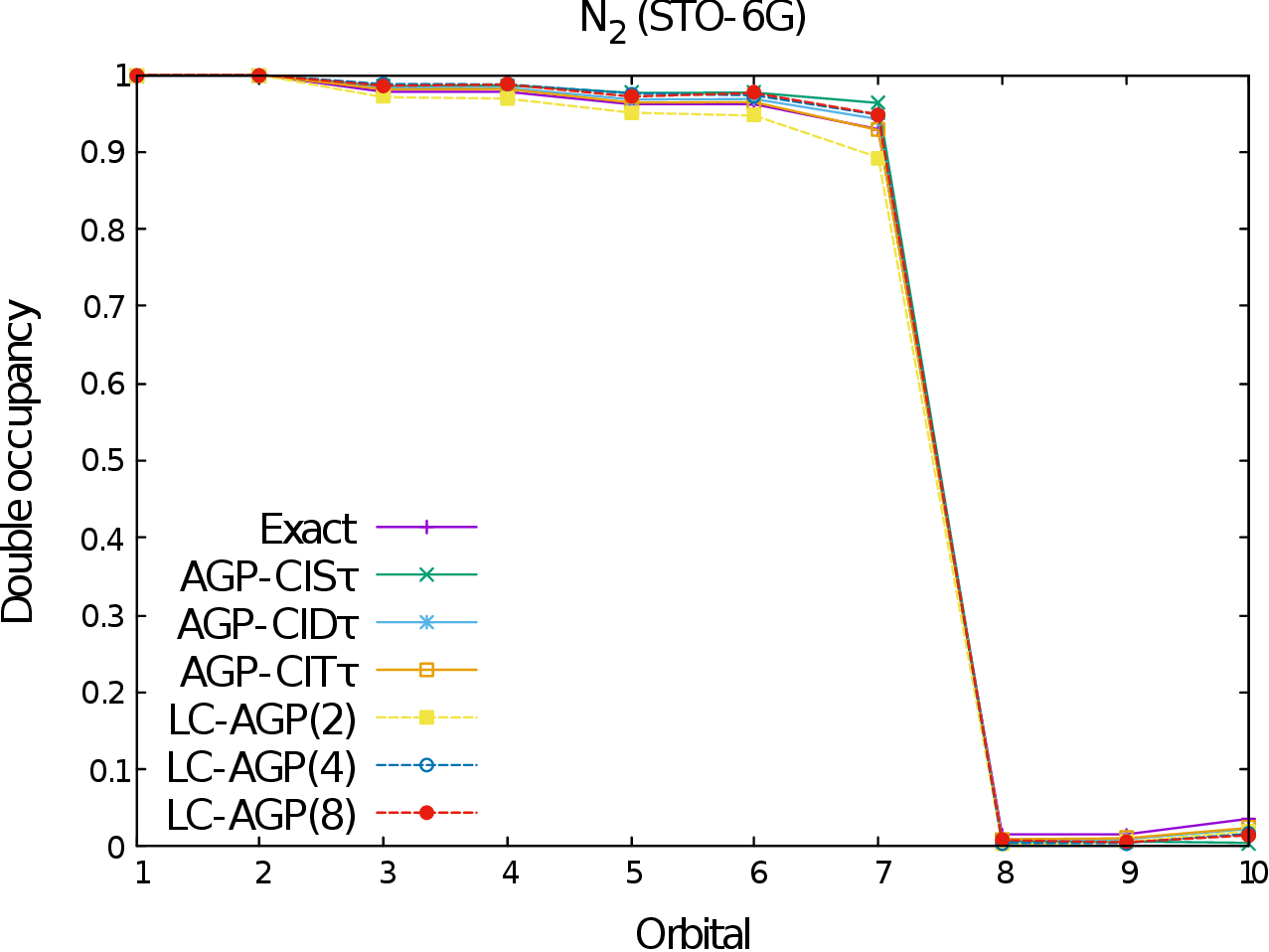}}}
\caption{Double occupancy for the N$_2$ (STO-6G, bond length = 2.3 bohr), orbitals 1–10.} 
\label{do_n2}
\end{figure}

In addition to the total energy, we analyzed the double occupancy, $\bra{\Psi}\hat{n}_{i,\uparrow} \hat{n}_{i,\downarrow}\ket{\Psi}$. Figure \ref{do_hubbard} shows the double occupancy for the 16-site, 16-electron Hubbard model at $U/t = 10$. Figure \ref{do_n2} shows the double occupancy for N$_2$ at a bond length of 2.3 bohr. Table \ref{rmse} summarizes the root-mean-square errors (RMSEs) of the double occupancy with respect to the exact results.
Hubbard model results show that, within LC-AGP, increasing the number of AGP terms does not lead to improved accuracy, and all tested LC-AGP wavefunctions perform worse than even AGP-CIS$\tau$.
For N$_2$, the behavior of the double occupancy closely follows the trends observed in the total energy. In particular, AGP-CIT$\tau$ yields values that are very close to the exact results.
Overall, these results indicate that, also for double occupancy, the AGP-CI$\tau$ wavefunctions provide better performance than LC-AGP.

\begin{table}
\caption{Root-mean-square error (RMSE) of the double occupancy for the Hubbard model (16-site, 16-electron system, $U/t = 10$) and N$_2$ (STO-6G, bond length = 2.3 bohr).}
{\begin{tabular}{l  cc} \hline
     & Hubbard model & N$_2$   \\ \hline
AGP-CIS$\tau$   & 0.005041 & 0.016941  \\
AGP-CID$\tau$   & 0.004666 & 0.007618  \\ 
AGP-CIT$\tau$   & 0.004409 & 0.004959  \\
LC-AGP ($K=2$)  & 0.005228 & 0.015197  \\
LC-AGP ($K=4$)  & 0.005217 & 0.012274  \\
LC-AGP ($K=8$)  & 0.005342 & 0.012105  \\ \hline
\end{tabular}}
\label{rmse}
\end{table}

\section{Conclusion} \label{s4}

In this paper, we developed a class of AGP-CI type wavefunctions to extend AGP by incorporating inter-geminal correlations through a CI expansion of AGP. In particular, we developed the AGP-CI$\tau$ wavefunctions, which employs a small parameter $\tau$ to suppress the growth of the number of AGP terms after polynomial decomposition.

The application of AGP-CI$\tau$ to the Hubbard model confirmed that, compared to LC-AGP, it maintains high accuracy as the number of electrons increases. In particular, for the strongly correlated case of $U/t=10$, LC-AGP shows a significant decrease in accuracy as the number of electrons increases, whereas AGP-CI$\tau$ maintains stable accuracy. Applications to small molecular systems also demonstrated that AGP-CI$\tau$ achieves high accuracy. In particular, for the strongly correlated N$_2$ molecule, LC-AGP does not show significant improvement even when the number of AGP terms is increased to 8, whereas AGP-CIT$\tau$ attains high accuracy.

LC-AGP faces the problem that the number of necessary AGP terms grows rapidly when attempting to improve accuracy by increasing electron number or considering strongly correlated systems. AGP-CI$\tau$, on the other hand, has the clear advantage that its accuracy increases systematically with the expansion of the CI terms. By extending the CI expansion in AGP-CI$\tau$, the wavefunction becomes nearly equivalent to that of APG. Therefore, AGP-CI$\tau$ is more flexible than APG and can reduce computational cost while maintaining the required accuracy.

A further aspect worth noting is the optimization behavior of LC-AGP. For instance, when comparing AGP-CIT$\tau$ with LC-AGP ($K=8$), one would expect LC-AGP to yield higher accuracy due to its larger variational space. However, it is likely that LC-AGP becomes trapped in local minima, preventing complete optimization and resulting in reduced accuracy. In fact, when the LC-AGP ($K=8$) calculation is performed using the optimized parameters from AGP-CIT$\tau$, a slightly lower total energy than that of AGP-CIT$\tau$ is obtained. The accuracy of LC-AGP may be improved, and further enhancement of AGP-CI$\tau$ could be possible, by refining the optimization procedure and the choice of initial parameters. Also, all molecular calculations in this work are carried out using the STO-6G basis set; however, further validation with larger basis sets, involving a greater number of variational parameters, will be required. Investigation of these possibilities will be the subject of future work.

\section*{Supplementary Material}

The supplementary material includes the total energy results for the Hubbard model at $U/t = 1$ and $-1$, and enlarged views of Figure \ref{do_n2}.

\begin{acknowledgments}

This work was supported by the U.S. Department of Energy, Office of Basic Energy Sciences, under Award DE-SC0001474.  G.E.S. is a Welch Foundation Chair (C-0036).
Some calculations were performed at the Research Center for Computer Science (RCCS), Institute for Molecular Science (IMS), Okazaki, Japan (Project: 25-IMS-C124). A.K. is grateful to Gunma University Graduate School of Science and Technology Overseas Research Visit for Young Faculty Members.

\end{acknowledgments}

\section*{Conflict of Interest}

The authors have no conflicts to disclose.

\section*{Data Availability Statement}

The data that support the findings of this study are available within the article.

\appendix

\section{AGP-CISD wavefunction} \label{app1}

AGP-CID wavefunction and AGP-CISD wavefunction can be expressed as follows.
\begin{eqnarray}
\ket{\Psi_{\mathrm{AGP-CID}} } &=& \hat{F}^n\ket{-}  + \hat{C}^{(1)}\hat{C}^{(2)}\hat{F}^{n-2} \ket{-}  \\
\ket{\Psi_{\mathrm{AGP-CISD}} } &=& \hat{F}^n\ket{-}  + \hat{C}^{(1)}\hat{F}^{n-1}\ket{-} + \hat{C}^{(1)}\hat{C}^{(2)}\hat{F}^{n-2} \ket{-} 
\end{eqnarray}
Here, by rearranging the AGP-CISD wavefunction, we obtain the following form.
\begin{eqnarray}
\ket{\Psi_{\mathrm{AGP-CISD}} } &=& \hat{F}^n\ket{-}  + \hat{C}^{(1)}\hat{F}^{n-1}\ket{-} + \hat{C}^{(1)}\hat{C}^{(2)}\hat{F}^{n-2} \ket{-} \nonumber \\
&=&\hat{F}^n\ket{-} + \hat{C}^{(1)} \left(\hat{F} + \hat{C}^{(2)}\right) \hat{F}^{n-2} \ket{-} \nonumber \\
&=& \hat{F}^n\ket{-}  + \hat{C}^{(1)}\hat{C}^{(2)'}\hat{F}^{n-2} \ket{-} \nonumber \\
&=& \ket{\Psi_{\mathrm{AGP-CID}} } \label{cisd}
\end{eqnarray}
where $\hat{C}^{(2)'}$ is defined as $\hat{C}^{(2)'} \equiv \hat{F} + \hat{C}^{(2)} $. Thus, one can see that the AGP-CISD wavefunction is equivalent to the AGP-CID wavefunction. Therefore, in the AGP-CI wavefunction, the lower-order terms in the expansion can be neglected. Note that Eq.~(\ref{cisd}) holds when the variational space of $C^{(2)}$ is unrestricted.

\section{AGP-CI$\tau$+AGP wavefunction} \label{app2}

The AGP-CI$\tau$ wavefunctions used in this paper differ from the AGP-CI wavefunctions in that the AGP term is not included. This approach is adopted to avoid numerical instability arising from the coexistence of terms with and without small values of $\tau$. When the AGP term is incorporated into AGP-CI$\tau$, the resulting wavefunctions are as follows.
\begin{eqnarray}
\ket{\Psi_{\mathrm{AGP-CIS}\tau+\mathrm{AGP}} } &=&  \hat{F}^n \ket{-} + \frac{1}{2n\tau} \left[ \left(\hat{F} + \tau \hat{C}^{(1)} \right)^n - \left(\hat{F} - \tau \hat{C}^{(1)}  \right)^n  \right] \ket{-}  \\
\ket{\Psi_{\mathrm{AGP-CID}\tau+\mathrm{AGP}} } &=& \hat{F}^n \ket{-}  \nonumber \\
&& + \frac{1}{4n(n-1)\tau^2} \left[ \left(\hat{F} + \tau( \hat{C}^{(1)} + \hat{C}^{(2)} ) \right)^n +\left(\hat{F} - \tau( \hat{C}^{(1)} + \hat{C}^{(2)} ) \right)^n  \right.  \nonumber \\
&& \hspace{2.2cm} \left. - \left(\hat{F} + \tau( \hat{C}^{(1)} - \hat{C}^{(2)} ) \right)^n- \left(\hat{F} - \tau( \hat{C}^{(1)} - \hat{C}^{(2)} ) \right)^n \right]  \ket{-}   \\
\ket{\Psi_{\mathrm{AGP-CIT}\tau+\mathrm{AGP}} } &=& \hat{F}^n \ket{-}  \nonumber \\
&& + \frac{1}{8n(n-1)(n-2)\tau^3} \left[ \left(\hat{F} + \tau( \hat{C}^{(1)} + \hat{C}^{(2)} + \hat{C}^{(3)}) \right)^n  +\left(\hat{F} - \tau( - \hat{C}^{(1)} + \hat{C}^{(2)} + \hat{C}^{(3)}  ) \right)^n  \right.  \nonumber \\
&&  \hspace{3.5cm}  + \left(\hat{F} - \tau( \hat{C}^{(1)} - \hat{C}^{(2)} + \hat{C}^{(3)}) \right)^n +\left(\hat{F} - \tau( \hat{C}^{(1)} + \hat{C}^{(2)} - \hat{C}^{(3)}  ) \right)^n \nonumber \\
&&  \hspace{3.5cm}  - \left(\hat{F} - \tau(  \hat{C}^{(1)} + \hat{C}^{(2)} + \hat{C}^{(3)}  ) \right)^n - \left(\hat{F} + \tau( -\hat{C}^{(1)} + \hat{C}^{(2)} + \hat{C}^{(3)}) \right)^n  \nonumber \\
&&  \hspace{3.5cm}  \left. - \left(\hat{F} + \tau(  \hat{C}^{(1)} - \hat{C}^{(2)} + \hat{C}^{(3)}  ) \right)^n - \left(\hat{F} + \tau( \hat{C}^{(1)} + \hat{C}^{(2)} - \hat{C}^{(3)}) \right)^n   \right] \ket{-}   \nonumber \\
\end{eqnarray}

\begin{figure}
\centering
{\resizebox*{10cm}{!}{\includegraphics{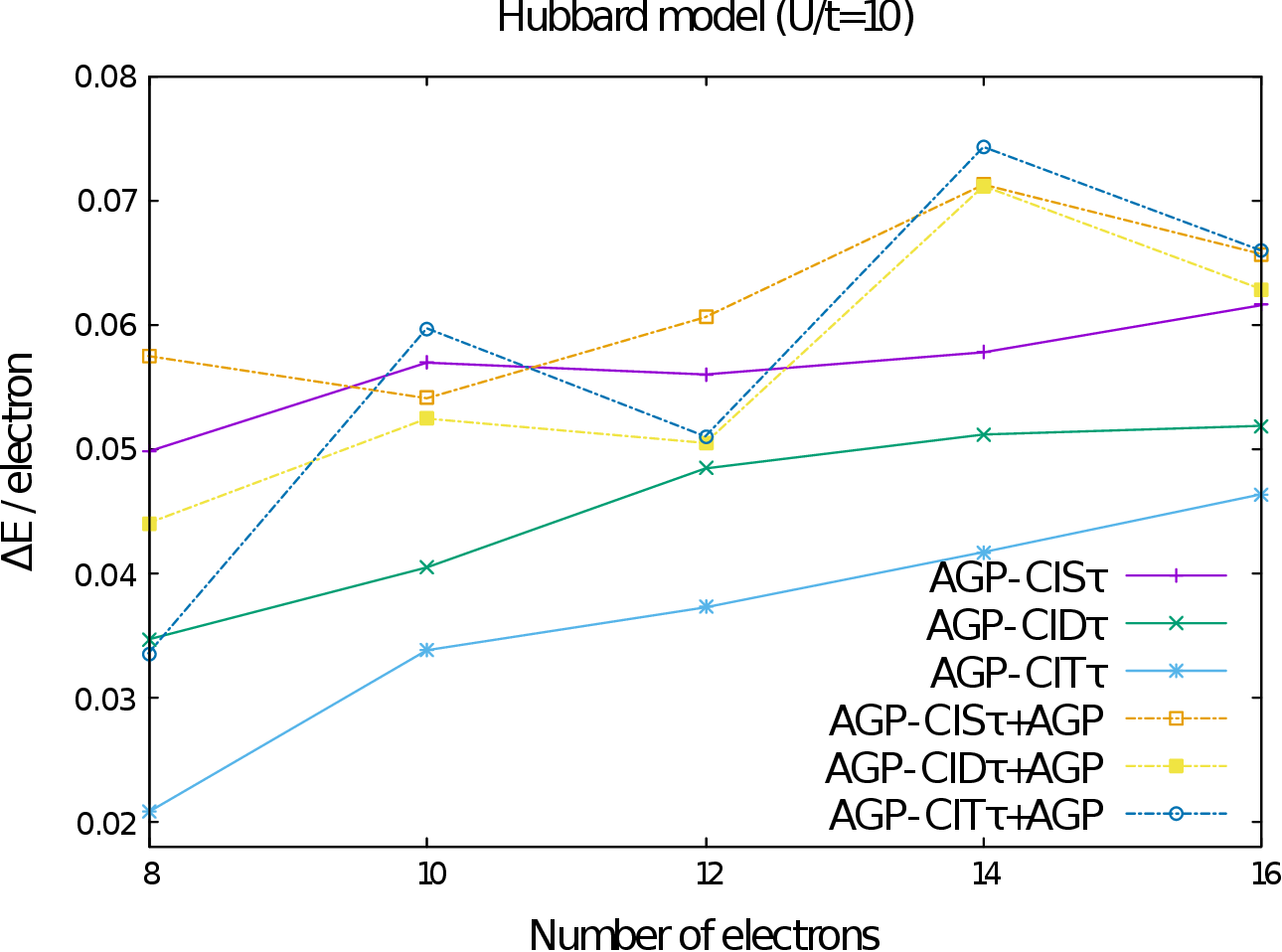}}}
\caption{Total energy error per electron for the Hubbard model ($U/t = 10$, half filling, $E/t$).} 
\label{energy_agp-cit+f}
\end{figure}

Table \ref{agp-cit+f} shows the total energy results of applying the AGP-CI$\tau$+AGP wavefunctions to the Hubbard model ($U/t=10$). Figure \ref{energy_agp-cit+f} shows the errors per electron relative to the exact solution, together with the AGP-CI$\tau$ wavefunctions for comparison. It can be seen that the AGP-CI$\tau$+AGP wavefunctions exhibit lower accuracy than the AGP-CI$\tau$ wavefunctions. In particular, the AGP-CIT$\tau$+AGP wavefunction, which incorporates higher-order CI expansions, is unstable and sometimes shows lower accuracy compared to AGP-CIS$\tau$+AGP. In this way, it can be seen that the AGP-CI$\tau$ wavefunctions are numerically more stable when the AGP term is not included. In addition, the computational cost is reduced by the amount corresponding to a single AGP term.

\begin{table}
\caption{Total energy for the Hubbard model ($U/t=10$, half filling, $E/t$).}
{\begin{tabular}{l  ccccc} \hline
Number of electrons    & 8 & 10 & 12 & 14 & 16   \\ \hline
Exact & -2.176688 & -2.703691 & -3.232383 & -3.762854 & -4.294284   \\  \hline
AGP-CIS$\tau$ + AGP  & -1.716973 & -2.162934 & -2.505191 & -2.765542 & -3.243880 \\
AGP-CID$\tau$ + AGP  & -1.824197 & -2.179114 & -2.626619 & -2.767095 & -3.288970 \\ 
AGP-CIT$\tau$ + AGP  & -1.907865 & -2.106650 & -2.620106 & -2.723141 & -3.239709 \\ \hline
\end{tabular}}
\label{agp-cit+f}
\end{table}

\section{$\tau$ as a variational parameter} \label{app3}

In this work, the AGP-CI$\tau$ wavefunctions are constructed with $\tau$ fixed at 0.2. Here, we consider the case in which $\tau$ is also treated as a variational parameter. To perform variational calculations, for example, the derivative of the AGP-CIS$\tau$ wavefunction with respect to $\tau$ is given as follows.
\begin{eqnarray}
\frac{\partial }{\partial \tau} \ket{\Psi_{\mathrm{AGP-CIS}\tau} } &=& -\frac{1}{2n\tau^2} \left[ \left(\hat{F} + \tau \hat{C}^{(1)} \right)^n - \left(\hat{F} - \tau \hat{C}^{(1)} \right)^n \right]\ket{-}  \nonumber \\
&& + \frac{1}{2\tau} \left[ \left(\hat{F} + \tau \hat{C}^{(1)} \right)^{n-1}\hat{C}^{(1)} + \left(\hat{F} - \tau \hat{C}^{(1)} \right)^{n-1}\hat{C}^{(1)} \right] \ket{-}  \nonumber \\
 &=& -\frac{1}{2n\tau^2} \left[ \left(\hat{F} + \tau \hat{C}^{(1)} \right)^n - \left(\hat{F} - \tau \hat{C}^{(1)} \right)^n \right]\ket{-}  \nonumber \\
 && + \frac{1}{2\tau} \left[ \sum_{k=0}^{n-1} \frac{1}{n} \frac{1}{(n-1-k)!k!}\frac{(-1)^{k}}{2^{n-1}}   \right.  \nonumber \\
 && \hspace{0.8cm}  \left. \times \left[\left(\hat{C}^{(1)} + (n-1-2k) \left(\hat{F} + \tau \hat{C}^{(1)}\right) \right)^{n} +\left( \hat{C}^{(1)} + (n-1-2k) \left(\hat{F} - \tau \hat{C}^{(1)} \right)\right)^{n} \right] \right] \ket{-}  \nonumber \\
\end{eqnarray}
Thus, the number of AGP terms arising from the derivative with respect to $\tau$ depends on $n$.

Table \ref{topt_table} shows the total energy results of applying AGP-CIS$\tau$ to the Hubbard model ($U/t=10$) for $\tau$ fixed at 0.2 and for $\tau$ optimized as a variational parameter. It can be seen that the accuracy of both approaches is almost identical. The slightly lower accuracy when $\tau$ is treated as a variational parameter is likely due to numerical instability caused by the increased number of variational parameters. Therefore, there is no need to increase the computational cost by treating $\tau$ as a variational parameter, and it is preferable to fix $\tau$ at a small value for the calculations.
\begin{table}
\caption{Total energy and optimized $\tau$ for the Hubbard model ($U/t=10$, half filling, $E/t$).}
{\begin{tabular}{l  cccccc} \hline
Number of electrons    & 8 & 10 & 12 & 14 & 16   \\ \hline
Exact & -2.176688 & -2.703690 & -3.232383 & -3.762854 &  -4.294284   \\  \hline
AGP-CIS$\tau$ ($t=0.2$)  & -1.778255 & -2.134351 & -2.560479 & -2.954145 & -3.309395  \\ \hline
AGP-CIS$\tau$ ($\tau=$ optimized $\tau$)  & -1.772435 & -2.128091 & -2.545311 & -2.918628 & -3.307764 \\ 
Optimized $\tau$                    &  0.158464 & -0.467412 & -0.567331 & -0.098061 & 0.224882  \\ \hline
\end{tabular}}
\label{topt_table}
\end{table}

\bibliography{cite}

\end{document}


\title{Supplementary material: Configuration interaction extension of AGP for incorporating inter-geminal correlations}
\author{Airi Kawasaki}
\email{a_kawasaki@gunma-u.ac.jp.}
\affiliation{Division of Electronics and Mechanical Engineering, Graduate School of Science and Technology, Gunma University, 1-5-1 Tenjin-cho, Kiryu-shi, Gunma 376-8515, Japan}
\author{Fei Gao}%
\affiliation{Department of Physics and Astronomy, Rice University, Houston, Texas 77005-1892, United States}%
\author{Gustavo E. Scuseria}
\affiliation{Department of Physics and Astronomy, Rice University, Houston, Texas 77005-1892, United States}%
\affiliation{Department of Chemistry, Rice University, Houston, Texas 77005-1892, United States}%

\date{\today}

\maketitle

\begin{figure}[H]
\centering
{\resizebox*{10cm}{!}{\includegraphics{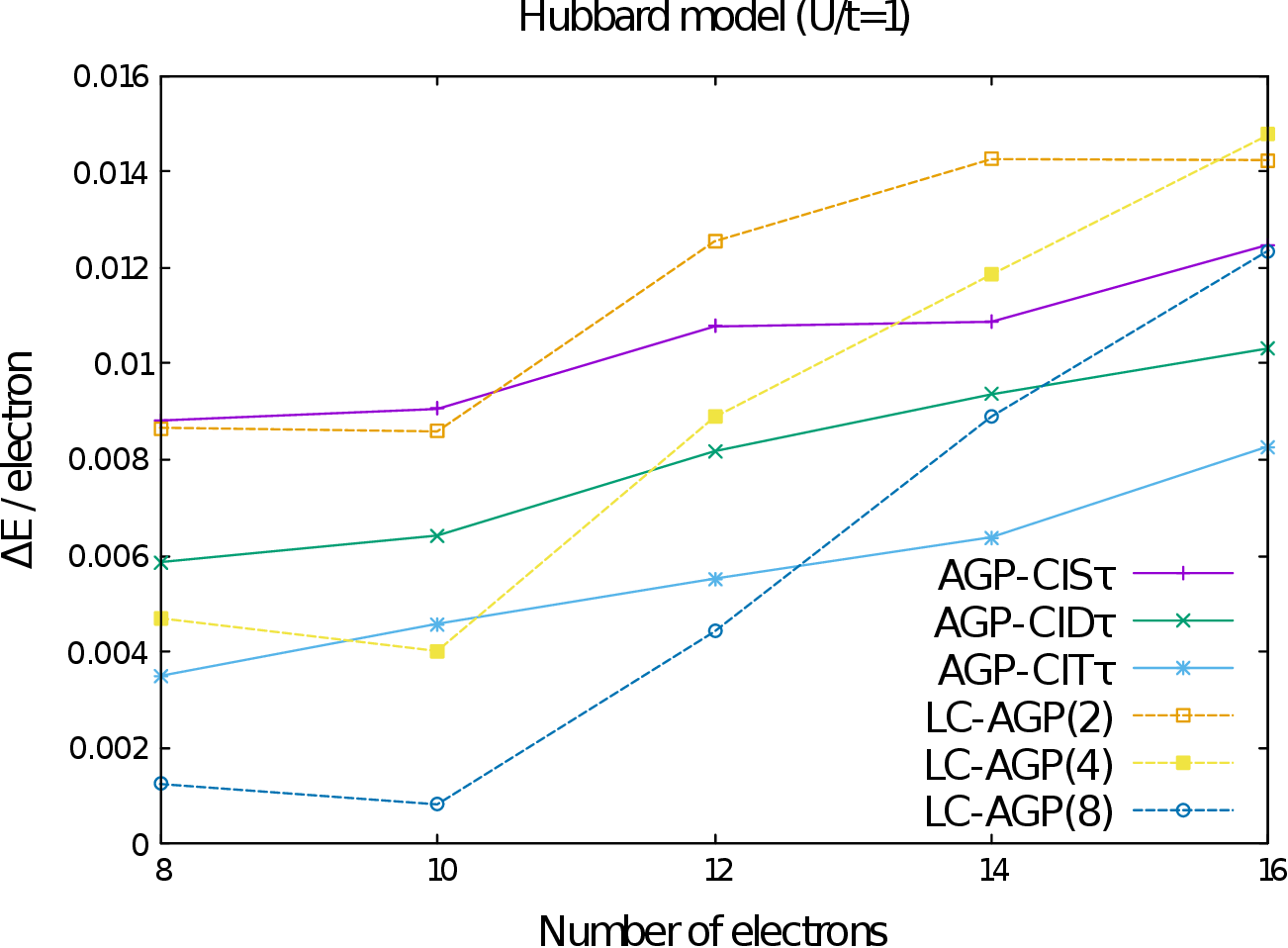}}}
\caption{Total energy error per electron for the Hubbard model ($U/t = 1$, half filling, $E/t$).} 
\label{energy_u1}
\end{figure}

\begin{figure}[H]
\centering
{\resizebox*{10cm}{!}{\includegraphics{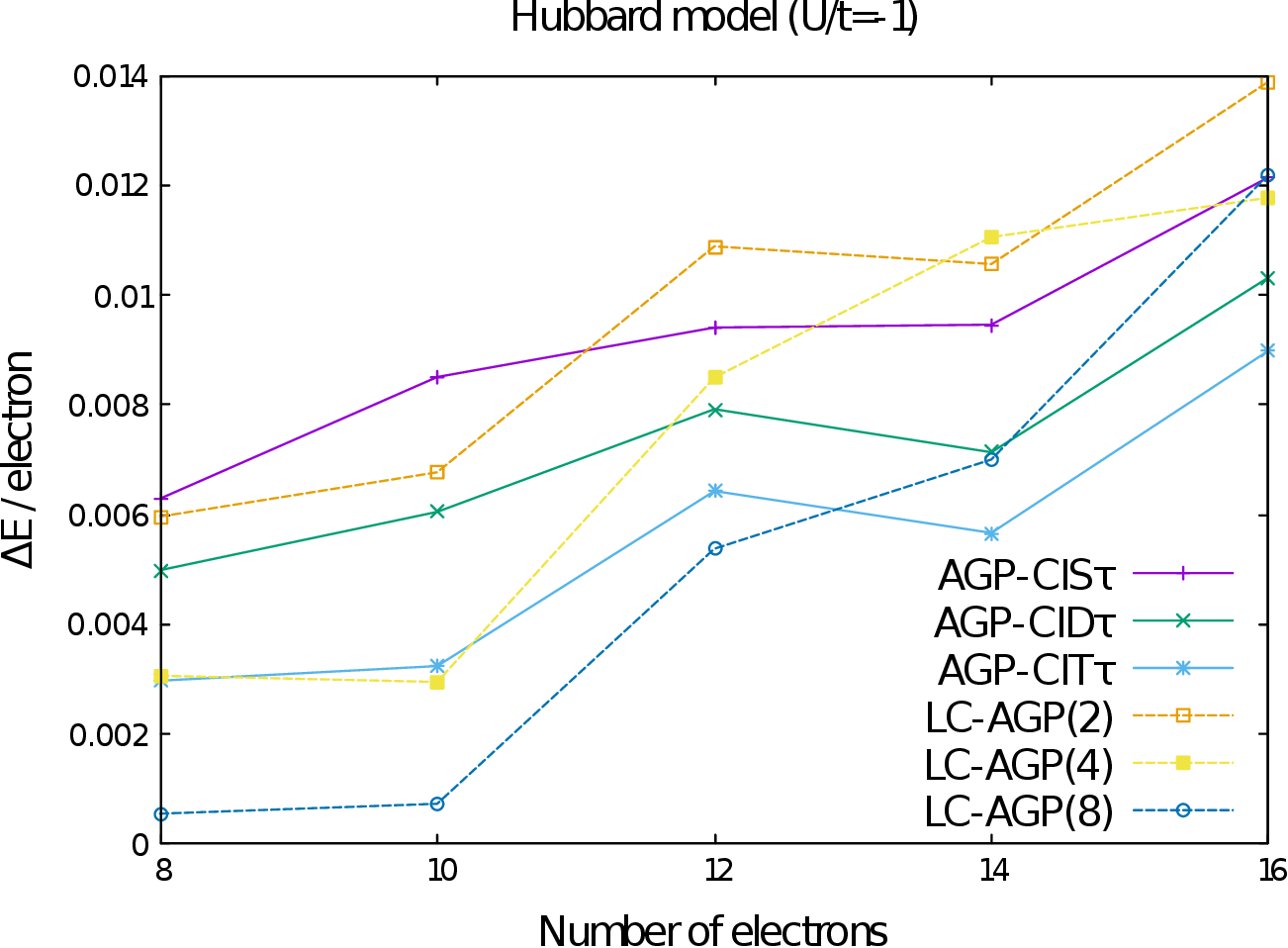}}}
\caption{Total energy error per electron for the Hubbard model ($U/t = -1$, half filling, $E/t$).} 
\label{energy_u-1}
\end{figure}

Tables \ref{hubbard_1} and \ref{hubbard_-1} show the total energy results obtained with AGP-CI$\tau$ and LC-AGP for the half-filled Hubbard model at $U/t = 1$ and $-1$ respectively. In addition, Figures \ref{energy_u1} and \ref{energy_u-1} show the errors per electron relative to the exact solution. AGP-CI$\tau$ retains its accuracy as the number of electrons increases, whereas LC-AGP fails to improve its accuracy with increasing numbers of AGP terms as the system size grows, exhibiting the same trend as seen for $U/t = 10$ in the main text.

\begin{table}[H]
\caption{Total energy for the Hubbard model ($U/t=1$, half filling, $E/t$).}
{\begin{tabular}{l  ccccc} \hline
Number of electrons    & 8 & 10 & 12 & 14 & 16   \\ \hline
Exact & -7.952326 & -10.61441 & -12.24929 & -14.71471 & -16.47587   \\  \hline
AGP-CIS$\tau$   & -7.881821 & -10.52387 & -12.12003 & -14.56253 & -16.27628 \\
AGP-CID$\tau$   & -7.905411 & -10.55021 & -12.15118 & -14.58371 & -16.31081 \\ 
AGP-CIT$\tau$   & -7.924342 & -10.56857 & -12.18301 & -14.62530 & -16.34343 \\
LC-AGP ($K=2$) & -7.883005 & -10.52855 & -12.09881 & -14.51517 & -16.24813 \\
LC-AGP ($K=4$) & -7.914736 & -10.57426 & -12.14270 & -14.54880 & -16.23969 \\
LC-AGP ($K=8$) & -7.942323 & -10.60617 & -12.19628 & -14.59021 & -16.27822 \\ \hline
\end{tabular}}
\label{hubbard_1}
\end{table}
\begin{table}[H]
\caption{Total energy for the Hubbard model ($U/t=-1$, half filling, $E/t$).}
{\begin{tabular}{l  ccccc} \hline
Number of electrons    & 8 & 10 & 12 & 14 & 16   \\ \hline
Exact & -11.95233 & -15.61441 & -18.24928 & -21.71471 & -24.47587   \\  \hline
AGP-CIS$\tau$   & -11.90201 & -15.52938 & -18.13642 & -21.58228 & -24.28160 \\
AGP-CID$\tau$   & -11.91244 & -15.55393 & -18.15428 & -21.61486 & -24.31078 \\ 
AGP-CIT$\tau$   & -11.92855 & -15.58201 & -18.17217 & -21.63537 & -24.33210 \\
LC-AGP ($K=2$) & -11.90463 & -15.54673 & -18.11868 & -21.56687 & -24.25372 \\
LC-AGP ($K=4$) & -11.92781 & -15.58490 & -18.14732 & -21.55996 & -24.28756 \\
LC-AGP ($K=8$) & -11.94791 & -15.60709 & -18.18476 & -21.61668 & -24.28076 \\ \hline
\end{tabular}}
\label{hubbard_-1}
\end{table}

\begin{figure}
\centering
{\resizebox*{10cm}{!}{\includegraphics{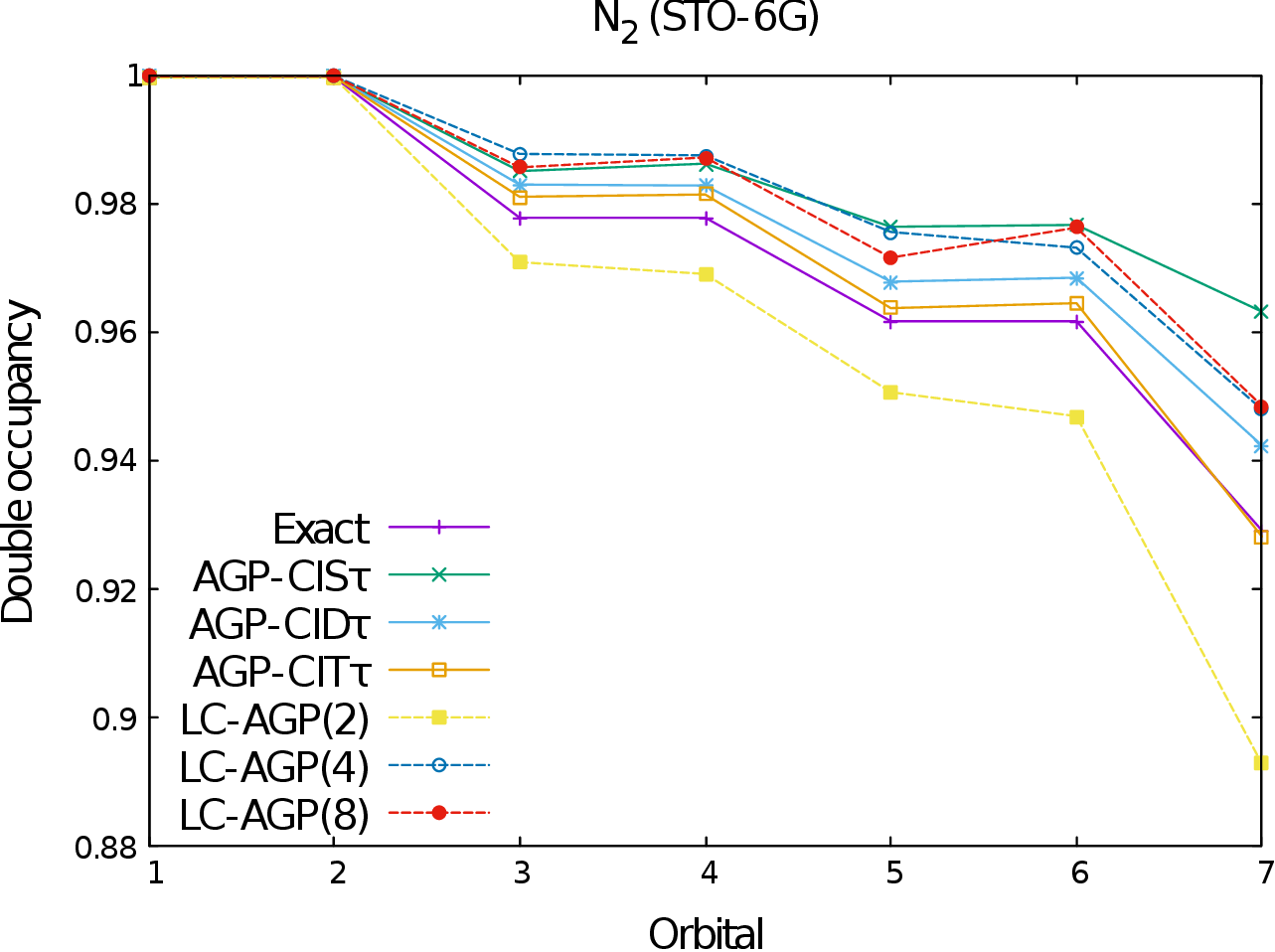}}}
\caption{Double occupancy for the N$_2$ (STO-6G, bond length = 2.3 bohr), orbitals 1–7.} 
\label{do_n2_1}
\end{figure}

\begin{figure}
\centering
{\resizebox*{10cm}{!}{\includegraphics{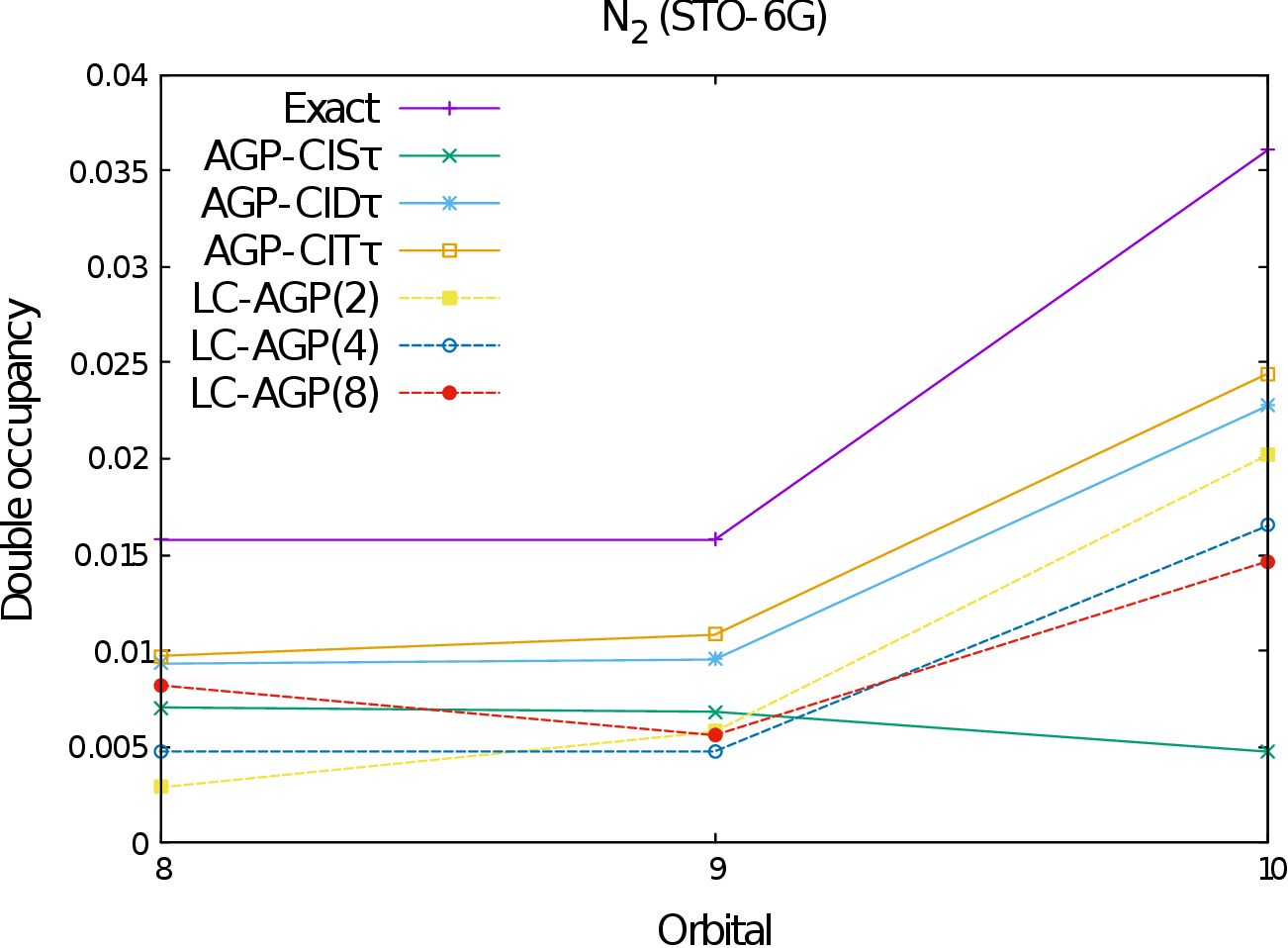}}}
\caption{Double occupancy for the N$_2$ (STO-6G, bond length = 2.3 bohr), orbitals 8–10.} 
\label{do_n2_2}
\end{figure}

Figures \ref{do_n2_1} and \ref{do_n2_2} provide enlarged views of the double occupancy for N$_2$ shown in the main text. It is evident from these figures that AGP-CIT$\tau$ yields results close to the exact solution.

